\documentclass[a4paper,12pt]{article}

\newcommand{\sect}[1]{\setcounter{equation}{0}\section{#1}}

\newcommand{\bea}{\begin{eqnarray}}
\newcommand{\ena}{\end{eqnarray}}
\renewcommand{\a}{\alpha}
\renewcommand{\b}{\beta}

\newcommand{\p}[1]{(\ref{#1})}

\begin{document}
\topmargin 0pt \oddsidemargin 0mm

\renewcommand{\thefootnote}{\fnsymbol{footnote}}

\begin{titlepage}

\begin{flushright}
CAS-KITPC/ITP-106\\
KU-TP 031 \\
\end{flushright}

\vspace{5mm}
\begin{center}
{\Large \bf Topological Black Holes in Ho\v{r}ava-Lifshitz Gravity} \vspace{15mm}

{\large Rong-Gen Cai$^{a,}$\footnote{e-mail address:
cairg@itp.ac.cn}, Li-Ming Cao$^{b,}$\footnote{e-mail address:
caolm@apctp.org}, Nobuyoshi Ohta$^{c,}$\footnote{e-mail address:
ohtan@phys.kindai.ac.jp}}

\vspace{10mm} {\em $^a$ Key Laboratory of Frontiers in Theoretical
Physics, Institute of Theoretical Physics, Chinese
Academy of Sciences,\\
P.O. Box 2735, Beijing 100190, China \\
Kavli Institute for Theoretical Physics China (KITPC), Chinese
Academy of Sciences,\\
P.O. Box 2735, Beijing 100190, China \\
$^{b}$Asia Pacific Center for
Theoretical Physics, Pohang,
Gyeongbuk 790-784, Korea \\
$^{c}$Department of Physics, Kinki University, Higashi-Osaka, Osaka
577-8502, Japan }

\end{center}

\vspace{10mm} \centerline{{\bf{Abstract}}} \vspace{5mm} We find
topological (charged) black holes whose horizon has an arbitrary
constant scalar curvature $2k$ in Ho\v{r}ava-Lifshitz theory.
Without loss of generality, one may take $k=1,0$ and $-1$. The black
hole solution is asymptotically AdS with a nonstandard asymptotic
behavior. Using the Hamiltonian approach, we define a finite mass
associated with the solution. We discuss the thermodynamics of the
topological black holes and find that the black hole entropy has a
logarithmic term in addition to an area term. We find a duality in
Hawking temperature between topological black holes in
Ho\v{r}ava-Lifshitz theory and Einstein's general relativity: the
temperature behaviors of black holes with $k=1, 0$ and $-1$ in
Ho\v{r}ava-Lifshitz theory are respectively dual to those of
topological black holes with $k=-1, 0$ and $1$ in Einstein's general
relativity. The topological black holes in Ho\v{r}ava-Lifshitz
theory are thermodynamically stable.

\end{titlepage}

\newpage
\renewcommand{\thefootnote}{\arabic{footnote}}
\setcounter{footnote}{0} \setcounter{page}{2}
\sect{Introduction}

Recently a field theory model for a UV complete theory of gravity
was proposed by Ho\v{r}ava~\cite{Hor1}, which is a non-relativistic
renormalisable theory of gravity and reduces to Einstein's general
relativity at large scales. This theory is named Ho\v{r}ava-Lifshitz
theory in the literature since at the UV fixed point of the theory
space and time have different scalings. Since then much attention
has been attracted to this gravity
theory~\cite{others,TS,Cal,KK,Klu,LMP,Muk,Brad}, including its
implications in cosmology~\cite{TS,Cal,KK,LMP,Muk,Brad}. In
\cite{LMP} the authors find some static spherically symmetric black
hole solutions in Ho\v{r}ava-Lifshitz theory.

In the $(3+1)$-dimensional ADM formalism, where the metric can be
written as
\begin{equation}
\label{eq1} ds^2=-N^2 dt^2 +g_{ij}(dx^i+N^idt)(dx^j+N^jdt),
\end{equation}
and for a spacelike hypersurface with a fixed time, its extrinsic
curvature $K_{ij}$ is
\begin{equation}
\label{eq2}
K_{ij}=\frac{1}{2N}(\dot g_{ij}-\nabla_iN_j-\nabla_jN_i),
\end{equation}
where a dot denotes a derivative with respect to $t$ and covariant
derivatives defined with respect to the spatial metric $g_{ij}$, the
action of  Ho\v{r}ava-Lifshitz theory is~\cite{Hor1}
\begin{eqnarray}
\label{eq3}
I &=& \int dt d^3x \sqrt{g}N \left ( \frac{2}{\kappa^2}
(K_{ij}K^{ij}-\lambda K^2) +\frac{\kappa^2\mu^2 (\Lambda
R-3\Lambda^2)}{8(1-3\lambda)}
+\frac{\kappa^2\mu^2(1-4\lambda)}{32(1-3\lambda)}R^2 \right. \nonumber \\
&& \left.
 -\frac{\kappa^2\mu^2}{8}R_{ij}R^{ij}+\frac{\kappa^2\mu}{2\omega^2}\epsilon^{ijk}R_{il}
\nabla_jR^l_{\ k}-\frac{\kappa^2}{2\omega^4}C_{ij}C^{ij} \right),
\end{eqnarray}
where $\kappa^2$, $\lambda$, $\mu$, $\omega$ and $\Lambda$ are
constant parameters and the Cotten tensor, $C_{ij}$, is defined by
\begin{equation}
\label{eq4}
C^{ij}=\epsilon^{ikl} \nabla_k \left (R^j_{\ l}-\frac{1}{4}R\delta^j_l\right)
= \epsilon^{ikl}\nabla_k R^j_{\ l} -\frac{1}{4}\epsilon^{ikj}\partial_kR.
\end{equation}
In (\ref{eq3}), the first two terms are the kinetic terms, while the
others give the potential of the theory in the so-called
``detailed-balance" form.

Comparing the action to that of general relativity, one can see that
the speed of light, Newton's constant and the cosmological constant
are
\begin{equation}
\label{eq5}
c=\frac{\kappa^2\mu}{4}\sqrt{\frac{\Lambda}{1-3\lambda}}, \ \
G=\frac{\kappa^2c}{32\pi }, \ \ \tilde \Lambda=\frac{3}{2}\Lambda,
\end{equation}
respectively. Let us notice that when $\lambda=1$, the first three
terms in (\ref{eq3}) could be reduced to the usual ones of
Einstein's general relativity. However, in Ho\v{r}ava-Lifshitz
theory, $\lambda$ is a dynamical coupling constant,
susceptible to quantum correction~\cite{Hor1}. In addition, we
see from (\ref{eq5}) that when $\lambda >1/3$, the cosmological
constant $\Lambda$ must be negative.
However, the cosmological constant can be positive if we make an analytic
continuation $\mu \to i\mu, w^2 \to -iw^2$~\cite{LMP}. In this paper, we consider
the former case with negative cosmological constant.

For later convenience, we rewrite the action~\p{eq3} as follows~\cite{LMP}:
\begin{eqnarray}
\label{eq6}
I &=& \int dtd^3x ({\cal L}_0 +{\cal L}_1), \\
 {\cal L}_0 &=& \sqrt{g}N \left \{\frac{2}{\kappa^2}
(K_{ij}K^{ij}-\lambda K^2) +\frac{\kappa^2\mu^2 (\Lambda
R-3\Lambda^2)}{8(1-3\lambda)}\right \},  \nonumber \\
 {\cal L}_1  &=& \sqrt{g}N \left \{\frac{\kappa^2\mu^2(1-4\lambda)}{32(1-3\lambda)}R^2
-\frac{\kappa^2}{2\omega^4}\left(C_{ij}-\frac{\mu
\omega^2}{2}R_{ij}\right)
\left(C^{ij}-\frac{\mu\omega^2}{2}R^{ij}\right) \right\}.\nonumber
\end{eqnarray}
The equations of motion for the action are given in \cite{KK,LMP},
but they are very lengthy and we will not reproduce them here.

In this note we are interested in black hole solutions in the action
(\ref{eq6}). Considering the static, spherically symmetric solutions
with the metric ansatz
\begin{equation}
\label{eq7}
ds^2 = -N^2(r) dt^2 +\frac{dr^2}{f(r)} +r^2 (d\theta^2 +\sin^2\theta d\phi^2).
\end{equation}
Without the term ${\cal L}_1$, the solution is the just the (A)dS
Schwarzschild black hole solution with metric functions~\cite{LMP}
\begin{equation}
\label{eq8}
N^2(r)=f(r)= 1-\frac{\Lambda}{2}r^2 -\frac{m}{r}.
\end{equation}
With the term ${\cal L}_1$, a general static, spherically
symmetric black hole solution with an arbitrary $\lambda$ is also
found in \cite{LMP}, but the solution is elusive.
We discuss the general solution in the appendix.
Of particular interest is the case with $\lambda=1$, on which we focus in the following.
The solution is then given by
\begin{equation}
\label{eq9}
N^2 =f = 1+x^2 -\alpha \sqrt{x},
\end{equation}
where $x=\sqrt{-\Lambda}r$ and $\alpha$ is an integration
constant. This solution is asymptotically $AdS_4$ and has a
singularity at $x=0$ if $\alpha \ne 0$. The singularity could be
covered by black hole horizon at $x_+$, the largest root of the
equation $f=0$ if $\alpha >0$. The Hawking temperature of the
black hole horizon is easily given by~\cite{LMP}
\begin{equation}
\label{eq10}
T= \frac{3x_+^2-1}{8\pi x_+ }\sqrt{-\Lambda}.
\end{equation}
Note that here we have corrected a typo in \cite{LMP}. One can see
from (\ref{eq10}) that there exists an extremal limit,
$x_+=1/\sqrt{3}$, where the temperature vanishes. Another remarkable
point one can see by comparing the solution (\ref{eq9}) and
(\ref{eq8}) is that general relativity is not always recovered at
large distance~\cite{LMP}. In addition, one may naively expect that
the mass of the black hole solution (\ref{eq9}) is divergent due to
the square root term.

The black hole solution (\ref{eq9}) is obtained from the action
(\ref{eq6}) in the detailed balance~\cite{Hor1}.  The authors in
\cite{LMP} also considered black hole solution in  Ho\v{r}ava-Lifshitz
theory without the condition of the detailed balance, namely in the theory given by
\begin{equation}
\label{eq11}
{\cal L} ={\cal L}_0 +(1-\epsilon^2){\cal L}_1,
\end{equation}
where $\epsilon$ is a constant. In this theory, the black hole
solution they found turns to be
\begin{equation}
\label{eq12}
N^2 =f = 1+\frac{x^2}{1-\epsilon^2}
-\frac{\sqrt{\alpha^2(1-\epsilon^2)x +\epsilon^2 x^4}}{1-\epsilon^2}.
\end{equation}
In the large distance limit, the solution reduces to
\begin{equation}
\label{eq13}
f =1 +\frac{x^2}{1+\epsilon}-\frac{\alpha^2}{2\epsilon x} + {\cal O}(x^{-4}).
\end{equation}
The authors in \cite{LMP} suggest that the solution has a finite
mass for non-vanishing $\epsilon$, while it becomes divergent as
$\epsilon=0$. In the latter case, the solution goes back to the one
(\ref{eq9}). Furthermore, when $\epsilon =1$, the solution becomes
the (A)dS Schwarzschild black hole solution~\p{eq8}.

In this note we are going to discuss thermodynamics of the black
hole solutions (\ref{eq9}) and (\ref{eq12}), which have not been
studied in \cite{LMP}. Since the solutions (\ref{eq9}) and
(\ref{eq12}) are asymptotically AdS, we will generalize those
solutions to the case of topological black holes with any constant
scalar curvature horizon~\cite{Topo,Cai,Cai04,CS}. We will also
discuss the topological charged black holes in Ho\v{r}ava-Lifshitz
theory by including Maxwell field.

\sect{Topological black holes and thermodynamics}

In this section we first generalize the spherically symmetric black
hole solution (\ref{eq9}) to the topological black hole case with
arbitrary constant scalar curvature horizon.  The black hole
solution is of the metric ansatz
\begin{equation}
\label{2eq1}
ds^2 =-\tilde N^2(r)f(r) dt^2 +\frac{dr^2}{f(r)} +r^2 d\Omega_k^2,
\end{equation}
where $d\Omega_k^2$ denotes the line element for an 2-dimensional
Einstein space with constant scalar curvature $2k$. Without loss of
generality, one may take $k=0$, $\pm 1$, respectively. Following
\cite{LMP}, substituting the metric (\ref{2eq1}) into (\ref{eq6}),
we find
\begin{eqnarray}
\label{2eq2}
I & =& \frac{\kappa^2\mu^2 \Lambda \Omega_k}{8(1-3\lambda)}\int dt dr
\tilde N \left \{-3\Lambda r^2 -2 (f-k) -2 r (f-k)' \right.
\nonumber \\
&&~~~~~~~~~~~~~~~~~~~~~~~ \left. +\frac{(\lambda-1)f'^2}{2\Lambda}
+\frac{(2\lambda-1)(f-k)^2}{\Lambda r^2} -\frac{2\lambda
 (f-k)}{\Lambda r}f'\right \},
\end{eqnarray}
where a prime denotes the derivative with respect to $r$ and
$\Omega_k$ is the volume of the 2-dimensional Einstein space.
Again, we consider the solution in the case of $\lambda=1$. In that
case, we have
\begin{equation}
\label{2eq3} I = \frac{\kappa^2\mu^2
\sqrt{-\Lambda}\Omega_k}{16}\int dt dx \tilde N \left( x^3-2x
(f-k)+\frac{(f-k)^2}{x}\right)'.
\end{equation}
Note that  here $x=\sqrt{-\Lambda}r$ and a prime becomes the
derivative with respect to $x$. From the action, we obtain the
equations of motion
\begin{eqnarray}
&& 0=\tilde N', \nonumber \\
&& 0=( x^3-2x (f-k)+\frac{(f-k)^2}{x})'.
\end{eqnarray}
From the first equation, we have $\tilde N=N_0$, a constant. One can
set $N_0=1$ by rescaling the time coordinate $t$. From the second
one, one can obtain $x^3-2x (f-k)+\frac{(f-k)^2}{x}=c_0$, here $c_0$
is an integration constant. Solving this yields
\begin{equation}
\label{2eq5} f(r)= k +x^2 -\sqrt{c_0 x}.
\end{equation}
Note that $c_0$ should be positive here. When $k=1$, the solution
reduces to the one given by \cite{LMP}. Thus we generalize the
solution in \cite{LMP} to the case of topological black holes with
arbitrary $k$. In addition, let us stress here that although we have
obtained the black hole solution through the minisuperspace
approach, it has been checked that the solution (\ref{2eq5}) with
$N_0=1$ indeed satisfies the equations of motion given in
\cite{LMP}.

A remarkable property of black holes is that they are associated
with thermodynamics. Now we are going to discuss thermodynamics of
the black hole solution (\ref{2eq5}), which has not yet been
discussed. Comparing to the AdS Schwarzschild black hole solution,
one may naively expect that the mass of the solution (\ref{2eq5}) is
divergent and one could not define a finite mass for this solution.
However, this conclusion is not true.  In fact, such non-standard
asymptotic behavior also appears for the black hole solutions in the
so-called dimensionally continued gravity~\cite{BTZ,CS}. For the
dimensionally continued black hole solutions, a finite mass can be
obtained by using the Hamiltonian approach. We find that this
approach also works for the Ho\v{r}ava-Lifshitz theory. Note that
the action (\ref{2eq3}) can be written as
\begin{equation}
\label{2eq6} I=\frac{\kappa^2\mu^2 \sqrt{-\Lambda}\Omega_k}{16}
(t_2-t_1)\int dx \tilde N \left( x^3-2x
(f-k)+\frac{(f-k)^2}{x}\right)' +B,
\end{equation}
where $B$ is a surface term, which must be chosen so that the
action has an extremum under variations of the fields with
appropriate boundary conditions. One demands that the fields
approach the classical solutions at infinity. Varying the action
(\ref{2eq6}), one finds the boundary term
\begin{equation}
\label{2eq7}
\delta B = -(t_2-t_1)N_0 \delta M.
\end{equation}
The boundary term $B$ is the conserved charge associated to the
``improper gauge transformations" produced by time
evolution~\cite{RT}. Here $M$ and $N_0$ are a conjugate pair.
Therefore when one varies $M$,  $N_0$ must be fixed. Thus the
boundary term should be in the form
\begin{equation}
\label{2eq8}
B =- (t_2-t_1) N_0 M  +B_0,
\end{equation}
where $B_0$ is an arbitrary constant, which should be fixed by
some physical consideration; for example, mass vanishes
when black hole horizon goes to zero. For details, see \cite{BTZ}.
According to this Hamiltonian approach, we get the mass of the
solution (\ref{2eq5}) as
\begin{equation}
\label{2eq9}
M= \frac{\kappa^2\mu^2 \sqrt{-\Lambda}\Omega_k}{16}c_0.
\end{equation}
Note that here $\Lambda$ is negative, therefore the black hole mass
is always positive because we have already set $c_0>0$. One can
easily obtain the Hawking temperature of the black hole, either by
directly calculating the surface gravity at the horizon, or by
requiring the absence of the conical singularity at the horizon of
the Euclidean black hole. Both methods give the same result
\begin{equation}
\label{2eq10}
T = \frac{3x_+^2 -k}{8\pi x_+} \sqrt{-\Lambda}.
\end{equation}
The next step is to get the entropy associated with the
topological black hole. In Einstein's general relativity, entropy of
black hole is always given by one quarter of black hole horizon
area.  But in higher derivative gravities, in general, the area
formula breaks down. Here we will obtain the black hole entropy by
using the first law of black hole thermodynamics with assumption
that as a thermodynamical system~\cite{Cai,Cai04,CS,BTZ}, the first
law always keeps valid: $ dM=TdS$. Integrating this relation yields
\begin{equation}
\label{2eq11}
S  \equiv \int T^{-1}dM +S_0 = \int T^{-1} \frac{dM}{dx_+}dx_+ +S_0,
\end{equation}
where $S_0$ is an integration constant, which should be fixed by
physical consideration. Through (\ref{2eq11}), we obtain
\begin{eqnarray}
\label{2eq12}
S &=& \frac{\pi \kappa^2 \mu^2 \Omega_k }{4}\left(x_+^2 +2 k\ln x_+
\right)
+S_0, \nonumber \\
&=& \frac{c^3}{4G}\left(A -\frac{k\Omega_k}{\Lambda} \ln
\frac{A}{A_0}\right),
\end{eqnarray}
where the Newton's constant and speed of light are given in
(\ref{eq5}), $A=\Omega_k r_+^2$ is the black hole horizon area, and
$A_0$ is a constant of dimension of length squared.  The leading
term is just one quarter of horizon area in units of $c=G=1$, which
should be the contribution from the ${\cal L}_0$ term. The second
term is a logarithmic function, therefore we cannot fix the
integration constant $S_0$ or $A_0$, unfortunately, by some physical
consideration, for example, black hole entropy should vanish when
black hole horizon goes to zero. The integration constant $S_0$
could be fixed by counting micro degrees of freedom in some quantum
theory of gravity like string theory. An interesting fact is that
such a term often appears in the quantum correction of black hole
entropy. In addition, when $k=0$, namely, for black hole with Ricci
flat horizon, the logarithmic term disappears. Thus, the area
formula of black hole entropy is recovered in this case. It might be
a universal result that the area formula still holds for Ricci flat
black holes in higher derivative gravity
theories~\cite{Cai,Cai04,CS}.

Two additional points are worth stressing here.
One is on the temperature (\ref{2eq10}).
For $k=1$, as pointed out in \cite{LMP},
there is an extremum at $x_+=1/\sqrt{3}$, where the temperature
vanishes, and it corresponds to an extremal black hole.
For $k=0$, the temperature $T=3x_+\sqrt{-\Lambda}/8\pi$.
In these two cases, the temperature always monotonically increases
as the horizon $x_+$ grows.
For $k=-1$, the inverse temperature starts from zero at $x_+=0$, monotonically
increases and reaches a maximal value,
$\beta=1/T=4\pi/\sqrt{-3\Lambda}$ at $x_+=1/\sqrt{3}$, then
monotonically decreases as $x_+$ grows. It is interesting to compare
these temperature behaviors of the topological black holes in
Ho\v{r}ava-Lifshitz theory with those for topological
black holes in Einstein's general relativity (the latter could be
obtained by replacing $1$ by $k$ in (\ref{eq8})). The temperature
for the topological black holes in Einstein's general relativity is
\begin{equation}
\label{2eq13}
T_{\rm TSch}= \frac{\sqrt{-\Lambda}}{8\pi x_+}(3x_+^2 +2k).
\end{equation}
We see that except for the coefficient difference in front of the
horizon curvature constant $k$, there is a duality relation in these
two temperatures: the temperature behaviors of black holes in
Ho\v{r}ova-Lifshitz theory in the cases of $k=1$,$0$ and $-1$, are
dual to the cases of $k=-1$, $0$ and $1$ in Einstein's general
relativity, respectively. Note that for topological black holes in
Einstein's general relativity~\cite{Topo}, in the cases of $k=0$ and
$k=-1$, the black holes are always thermodynamically stable, while
in the case of $k=1$, the small black hole with $x_+ <\sqrt{2/3}$ is
thermodynamically unstable and it becomes thermodynamically stable
for large horizon radius $x_+>\sqrt{2/3}$.

However, a close check tells us that in the case of $k=-1$, there
exists a minimal horizon at $x_+=1$ for the topological black hole
in Ho\v{r}ava-Lifshitz theory, which can be seen from
the metric function $f(r)$ in (\ref{2eq5}), namely for the case of
$c_0=0$.  This is just the massless black hole in AdS space. Thus in
the range $x_+ \in [1, \infty)$, the temperature of the topological
black hole is also a monotonically increasing function of $x_+$.
Thus the unstable phase for the topological black hole with $k=-1$
in Ho\v{r}ava-Lifshitz theory does not appear, and the black hole is
always thermodynamically stable.

To see this more clearly, let us calculate heat capacity of black hole,
defined as $C=dM/dT$.
The heat capacity of the black hole in Ho\v{r}ava-Lifshitz gravity is
\begin{equation}
\label{2eq14}
C= \frac{\pi \kappa^2\mu^2\Omega_k}{2}
\frac{(3x_+^2-k)(x_+^2+k)}{3x_+^2+k}.
\end{equation}
We see that for the cases $k=1$ and $k=0$, the heat capacity is
always positive, which implies that the black hole is local
thermodynamically stable, while in the case of $k=-1$, if $x_+>1$,
it is also positive. For comparison, we give the heat capacity for the
topological AdS black hole in Einstein's general relativity
\begin{equation}
\label{2eq15}
C_{\rm TSch} =\frac{\pi \kappa^2\mu^2 \Omega_k}{2}
\frac{3x_+^2+2k}{3x_+^2-2k}x_+^2.
\end{equation}
When $k=0$ and $-1$, it is always positive while when $k=1$, it is
negative for $x_+^2<2/3$, positive for $x_+^2>2/3$ and diverges at
$x_+^2=2/3$.

Another interesting question is whether there exists the Hawking-Page phase
transition associated with the black holes in Ho\v{r}ava-Lifshitz gravity.
It is well known that there is a Hawking-Page transition for static, spherically
symmetric AdS-Schwarzschild black hole (the case of $k=1$) between a
large  AdS black hole  and thermal gas in AdS space~\cite{HP}. On
the other hand, for the cases of $k=0$ and $k=-1$ topological black
hole in Einstein's general relativity, the Hawking-Page phase
transition does not exist. To discuss the Hawking-Page transition,
one has to calculate the Euclidean action or free energy of the
black hole. The Euclidean action has a relation to the free energy
by $I =\beta F$, here $\beta$ is the inverse temperature of the
black hole. By definition, the free energy $F$ is given by $F=M-TS$.
By using (\ref{2eq9}), (\ref{2eq10}) and (\ref{2eq12}), we find
\begin{equation}
\label{2eq16}
F= \frac{\kappa^2\mu^2\Omega_k\sqrt{-\Lambda}}{32x_+}
\left( -x_+^4 +5kx_+^2 +2k^2-6kx_+^2\ln x_+ +2k^2\ln x_+\right) -TS_0.
\end{equation}
Due to the uncertainty of $S_0$, we cannot determine the signature
of the free energy. However, if one can neglect the term $S_0$, we
see the free energy is negative for large enough horizon radius,
which means that large black holes in Ho\v{r}ava-Lifshitz gravity is
thermodynamically stable globally.

Now we turn to the case without the detailed balance condition,
namely $\epsilon^2 \ne 0$. Replacing (\ref{2eq3}) we have
\begin{equation}
\label{2eq17} I = \frac{\kappa^2\mu^2
\sqrt{-\Lambda}\Omega_k}{16}\int dt dx \tilde N \left( x^3-2x
(f-k)+(1-\epsilon^2)\frac{(f-k)^2}{x}\right)'.
\end{equation}
In this case, one has the solution
\begin{eqnarray}
\label{2eq18}
&& \tilde N=N_0, \nonumber \\
&& f(r)= k+ \frac{x^2}{1-\epsilon^2}
-\frac{\sqrt{\epsilon^2x^4+(1-\epsilon^2)c_0x}}{1-\epsilon^2}.
\end{eqnarray}
Again, $c_0$ is an integration constant and $N_0$ could be set
to one. Similar to the case of $\epsilon^2=0$, we find the mass
of the solution is
\begin{equation}
\label{2eq19}
M= \frac{\kappa^2\mu^2 \Omega_k\sqrt{-\Lambda}}{16}c_0,
\end{equation}
and $c_0$ can be expressed in terms of black hole horizon radius
$x_+$,
\begin{equation}
\label{2eq20}
c_0= \frac{x_+^4+2kx_+ +(1-\epsilon^2)k^2}{x_+}.
\end{equation}
The Hawking temperature of the black hole is found to be
\begin{equation}
\label{2eq21}
T = \frac{\sqrt{-\Lambda}}{8\pi}
\frac{3x_+^4+2kx_+^2-(1-\epsilon^2)k^2}{x_+(x_+^2+(1-\epsilon^2)k)}.
\end{equation}
With the mass and temperature, we obtain the entropy of the black hole
\begin{eqnarray}
\label{2eq22}
S &=&\frac{\pi \kappa^2 \mu^2 \Omega_k }{4}\left(x_+^2 +2 k(1-\epsilon^2)\ln x_+
\right) +S_0, \nonumber \\
&=& \frac{c^3}{4G}\left(A -(1-\epsilon^2)\frac{k\Omega_k}{\Lambda}
\ln \frac{A}{A_0}\right).
\end{eqnarray}
When $\epsilon^2=0$, it goes back to (\ref{2eq12}), while it reduces
to the well-known area formula for $\epsilon^2=1$, as expected, since
in that case, the effect of higher derivative terms disappears.

Now let us discuss the behavior of the temperature (\ref{2eq21}).

(i) When $k=0$, the temperature is independent of $\epsilon^2$,
given by
\begin{equation}
\label{2eq23}
T= \frac{3\sqrt{-\Lambda}}{8\pi}x_+.
\end{equation}
Clearly it is a monotonically increasing function of $x_+$

(ii)  When $k=-1$ and $\epsilon^2<1$, an extremal black hole with $T=0$ is obtained
at $x_ +^2= (1+\sqrt{1+3(1-\epsilon^2)})/3$. While to keep the denominator in
(\ref{2eq21}) positive, one has to have $x_+^2 >(1-\epsilon^2)$,
which is always smaller than $(1+\sqrt{1+3(1-\epsilon^2)})/3$.
This indicates that there does exist an extremal black hole in this case
with the minimal horizon radius $x_{+\rm min}^2=(1+\sqrt{1+3(1-\epsilon^2)})/3$.
When $\epsilon^2>1$, according to (\ref{2eq18}), the
minimal horizon radius is $x_+^2= 1+\epsilon$. In both cases of
$\epsilon^2>1$ and $<1$, the temperature of the black hole is a
monotonically increasing  function of $x_+$ in the physical regime.

(iii) When $k=1$, let us first consider the case of
$\epsilon^2<1$. A vanishing temperature happens at
$x^2_{+\rm min}= (-1+\sqrt{1+3(1-\epsilon^2)})/3$. When
$\epsilon^2>1$, there does not exist an extremal black hole, but
keep the temperature positive, a physical horizon radius must obey
$x_+^2 >\epsilon^2-1$. As the case of $k=-1$ with any $\epsilon^2$,
the temperature of the black hole is a
monotonically increasing  function of $x_+$ in the physical regime, again.

In summary, the case with $\epsilon^2 \ne 0$ is similar to the case
with $\epsilon^2=0$, the Hawking temperature of the black holes with
any $k$ is always a monotonically increasing function of horizon
radius $x_+$ in the physical regime. This implies that the
topological black holes in Ho\v{r}ava-Lifshitz theory are
thermodynamically stable. Note that when $\epsilon^2=1$, the
situation is reduced to the case of the well-known topological AdS
Schwarzschild black holes~\cite{Topo}.

\sect{Topological charged black holes}

In this section we consider the charged generalization of the
topological black hole found in Sec.~2.  To give a universal
result, we assume $\epsilon^2 \ne 0$.  Following \cite{BTZ,CS},
the Hamiltonian action for the Maxwell field can be written as
\begin{equation}
\label{4eq1}
I_{\rm em} =\int dt d^3x \left [ p^i\dot
A_i-\frac{1}{2}N\left(\alpha
g^{-1/2}p^ip_i+\frac{g^{1/2}}{2\alpha} F_{ij}F^{ij}\right)
+\varphi p^i,_i\right] +B_{\rm em},
\end{equation}
where $p^i$ is the momentum conjugate of the spatial components
of the Maxwell field $A_i$, $\varphi =A_0$, $B_{\rm em}$ is a
boundary term, $N$ is the lapse function, and $\alpha$ is a parameter to
be fixed shortly. Considering the static topological black
hole solution with the metric ansatz (\ref{2eq1}), the action
(\ref{4eq1}) is reduced to
\begin{equation}
\label{4eq2}
I_{\rm em} = \frac{\Omega_k}{\alpha} \int dt dr
\left(-\frac{1}{2}\tilde N r^2 p^2 +\varphi (r^2p)'\right) +B_{\rm em},
\end{equation}
where $p= \alpha p^r/r^2\gamma^{1/2}$ and $\gamma$ is the
determinate of the 2-dimensional Einstein space $d\Omega^2_k$.
Note that here the solution without magnetic charge $F_{ij}=0$ has
been assumed. To be consistent with (\ref{2eq17}), we set
$x=\sqrt{-\Lambda}r$, The action (\ref{4eq2}) then becomes
\begin{equation}
\label{4eq3}
I_{\rm em}= \frac{\Omega_k}{\alpha \sqrt{-\Lambda}}\int dt dx
\left(-\frac{1}{2}\tilde N x^2 \tilde p^2 +\varphi (x^2\tilde p)'\right) +B_{\rm em},
\end{equation}
where a prime denotes derivative with respective to $x$ and
$\tilde p = p/\sqrt{-\Lambda}$. Now we set
\begin{equation}
\label{4eq4}
\alpha^{-1} = -\frac{\kappa^2\mu^2 \Lambda}{16}.
\end{equation}
Combining (\ref{2eq17}) and (\ref{4eq4}), we have
\begin{equation}
\label{4eq5}
I =\frac{\kappa^2\mu^2
\sqrt{-\Lambda}\Omega_k}{16}\int dt dx \left(\tilde N
(U'-\frac{1}{2}x^2\tilde p^2) +\varphi (x^2\tilde p)'\right) +B,
\end{equation}
where
$$ U =x^3-2x
(f-k)+(1-\epsilon^2)\frac{(f-k)^2}{x}.$$
{}From the action (\ref{4eq5}) we obtain the equations of motion
\begin{eqnarray}
\label{4eq6}
&& U'= \frac{1}{2} x^2 \tilde p^2, \ \ \
(x^2 \tilde p)' =0, \nonumber \\
&& \varphi' = -\tilde N \tilde p, \ \ \
\tilde N'=0,
\end{eqnarray}
which have the solution
\begin{eqnarray}
\label{4eq7}
&& \tilde N =N_0, \ \ \ \ \varphi = \frac{N_0 q}{x} +\varphi_0, \nonumber \\
&& \tilde p = \frac{q}{x^2}, \ \ \ \
U = -\frac{q^2}{2x}+c_0.
\end{eqnarray}
Here $N_0$, $\varphi_0$, $c_0$ and $q$ are integration constants,
their physical meanings are clear. Physical electric charge and
mass of the solution are
\begin{equation}
\label{4eq8}
Q = \frac{\kappa^2 \mu^2 \Omega_k\sqrt{-\Lambda}}{16}q, \ \ \  M= \frac{\kappa^2 \mu^2
\Omega_k\sqrt{-\Lambda}}{16}c_0,
\end{equation}
respectively, and the metric function $f$ is given by
\begin{equation}
f(r) =  k +\frac{x^2}{1-\epsilon^2} - \frac{\sqrt{\epsilon^2x^4
+(1-\epsilon^2)(c_0x-q^2/2)}}{1-\epsilon^2},
\end{equation}
while $\tilde N=N_0 $ could be set to one.  Taking the limit $\epsilon \to 1$,
the solution is reduced to
\begin{equation}
f(r) = k +\frac{x^2}{2} -\frac{c_0}{2x} +\frac{q^2}{4x^2},
\end{equation}
as expected, it is just the AdS Reissner-Nordstr\"om black hole
solution. The Hawking temperature of the black hole is
\begin{equation}
\label{4eq11}
T =\frac{\sqrt{-\Lambda} (3x_+^4 +2kx_+^2 -(1-\epsilon^2)k^2 -q^2/2)}{8\pi x_+
(x_+^2+(1-\epsilon^2)k)}.
\end{equation}
Putting the temperature (\ref{4eq11}) and mass (\ref{4eq8}) into the
first law of black hole thermodynamics, it is easy to check that one
reproduces the entropy (\ref{2eq22}), the charge $q$ does not
appear explicitly in the expression of black hole entropy in terms
of horizon radius. This is consistent with the fact that black hole
entropy is a function of horizon geometry. The behavior of the
temperature can be analyzed as the case without the electric charge,
but we do not repeat here.  Instead we only point out that due to
the appearance of the electric charge, extremal black holes with
vanishing temperature always exist within reasonable parameter
regime.

\sect{Conclusion}

In this paper we found topological (charge) black hole solutions
with arbitrary constant scalar curvature horizon in
Ho\v{r}ava-Lifshitz theory, generalizing the static, spherically
symmetric black hole solutions in \cite{LMP}. Although there is a
square root term in the metric function $f(r)$, we can define a
finite mass associated with the black hole solution by use of the
Hamiltonian approach. We have calculated the Hawking temperature of
the black hole and the black hole entropy by using the first law
of black hole thermodynamics, and found that, except for the
well-known horizon area term, the black hole entropy has a
logarithmic term. Such a logarithmic term often occurs on the
occasion of considering quantum corrections to black hole entropy.
In our entropy expression, there is a undetermined constant $S_0$.
To fix the constant entropy $S_0$, one has to invoke quantum theory
of gravity.

We find that the temperature behavior of the topological black holes
in Ho\v{r}ava-Lifshitz theory is very interesting. Indeed there is a
duality for temperature between topological black holes in
Ho\v{r}ava-Lifshitz theory and topological black holes in Einstein's
general relativity. The temperatures of topological black holes with
$k=1$, $0$ and $-1$ in Ho\v{r}ava-Lifshitz theory are dual to those
of black holes with $k=-1$, $0$ and $1$ in Einstein's general
relativity, respectively.

In this paper we have only considered  thermodynamics of
topological black holes in Ho\v{r}ava-Lifshitz theory
with $\lambda=1$. It is of great interest to see whether one can
find a way to study thermodynamics for the general topological
black holes in the theory with $\lambda \ne 1$.

\section*{Acknowledgments}
This work was supported partially by grants from NSFC, China (No.
10821504 and No. 10525060), a grant from the Chinese Academy of
Sciences with No.KJCX3-SYW-N2, the Grant-in-Aid for
Scientific Research Fund of the JSPS No. 20540283, and
the Japan-U.K. Research Cooperative Program.

\appendix
\section{Appendix: Topological black holes for general $\lambda$}

Here we briefly discuss topological black hole solution with a
general $\lambda$. In terms of the new function $F$ defined by \bea
F(r)=k-\Lambda r^2 -f(r), \label{3eq1} \ena the action~\p{2eq2}
takes the form \bea I = \frac{\kappa^2\mu^2
\Omega_k}{8(1-3\lambda)}\int dt dr \tilde N \left \{
\frac{(\lambda-1)}{2} F'^2
 - \frac{2\lambda}{r} FF' +\frac{(2\lambda-1)}{r^2} F^2 \right \}.
\label{3eq2}
\ena
The equations of motion are then
\bea
\label{3eq31}
&& 0= \left(\frac{2\lambda}{r}F-(\lambda-1)F' \right) \tilde N'
+(\lambda-1)\left(\frac{2}{r^2}F-F'' \right)\tilde N, \\
&& 0= (\lambda-1)r^2 F'^2 -4\lambda rFF'+2(2\lambda-1)F^2.
\label{3eq32} \ena The latter is easily solved to give~\cite{LMP}
\bea F(r)= \a r^{\frac{2\lambda\pm\sqrt{2(3\lambda-1)}}{\lambda-1}},
\label{3eq4} \ena and then the first gives \bea \tilde N = \b
r^{-\frac{1+3\lambda \pm 2\sqrt{2(3\lambda-1)}}{\lambda-1}},
\label{3eq5} \ena where $\a$ and $\b$ are both integration
constants. When $\a=0$ or $F=0$, Eq.~\p{3eq31} does not restrict
$\tilde N$. Note that the exponent of Eq.~\p{3eq4} for the negative
branch is always less than 2 for positive $\lambda$, and thus the
$r^2$ term in the metric function~\p{3eq1} dominates at large
distances. The other branch gives a power larger than 2. We are
interested in the solutions with asymptotic AdS behavior. In that
case, we should look at the negative branch with constant $\tilde
N$. It follows from Eq.~\p{3eq31} that either $\lambda=1$ or
$F''=\frac{2}{r^2}F$. The latter leads to $F \sim r^2$ or $1/r$; the
first one does not satisfy \p{3eq32}, and the second solution
requires $\lambda=1/3$, which may be of some interest~\cite{Hor1},
but the action~\p{eq3} appears singular. So we discuss $\lambda=1$
case mainly in this paper.



\begin{thebibliography}{99}
\bibitem{Hor1} P.~Horava,
  arXiv:0901.3775 [hep-th].
\bibitem{others} P.~Horava,
  JHEP {\bf 0903}, 020 (2009)
  [arXiv:0812.4287 [hep-th]];
   M.~Visser,
  arXiv:0902.0590 [hep-th];
  L.~Maccione, A.~M.~Taylor, D.~M.~Mattingly and S.~Liberati,
  arXiv:0902.1756 [astro-ph.HE];
    P.~R.~S.~Carvalho and M.~M.~Leite,
  arXiv:0902.1972 [hep-th];
   P.~Horava,
  arXiv:0902.3657 [hep-th];
   A.~Volovich and C.~Wen,
  arXiv:0903.2455 [hep-th];
   A.~Jenkins,
  arXiv:0904.0453 [gr-qc].

\bibitem{TS}  T.~Takahashi and J.~Soda,
  arXiv:0904.0554 [hep-th].

\bibitem{Cal} G.~Calcagni,
  arXiv:0904.0829 [hep-th].
\bibitem{KK} E.~Kiritsis and G.~Kofinas,
  arXiv:0904.1334 [hep-th].
\bibitem{Klu} J.~Kluson,
  arXiv:0904.1343 [hep-th].
\bibitem{LMP}  H.~Lu, J.~Mei and C.~N.~Pope,
  arXiv:0904.1595 [hep-th].
\bibitem{Muk} S.~Mukohyama,
  arXiv:0904.2190 [hep-th].

\bibitem{Brad} R.~Brandenberger,
  arXiv:0904.2835 [hep-th].

\bibitem{Topo}J.~P.~S.~Lemos,
  Phys.\ Lett.\  B {\bf 353}, 46 (1995)
  [arXiv:gr-qc/9404041];
  J.~P.~S.~Lemos and V.~T.~Zanchin,
  Phys.\ Rev.\  D {\bf 54}, 3840 (1996)
  [arXiv:hep-th/9511188];
  C.~G.~Huang and C.~B.~Liang,
  Phys.\ Lett.\  A {\bf 201}, 27 (1995);
  R.~G.~Cai and Y.~Z.~Zhang,
  Phys.\ Rev.\  D {\bf 54}, 4891 (1996)
  [arXiv:gr-qc/9609065];
  S.~Aminneborg, I.~Bengtsson, S.~Holst and P.~Peldan,
  Class.\ Quant.\ Grav.\  {\bf 13}, 2707 (1996)
  [arXiv:gr-qc/9604005];
  R.~B.~Mann,
  Class.\ Quant.\ Grav.\  {\bf 14}, L109 (1997)
  [arXiv:gr-qc/9607071];
  D.~R.~Brill, J.~Louko and P.~Peldan,
  Phys.\ Rev.\  D {\bf 56}, 3600 (1997)
  [arXiv:gr-qc/9705012].
  L.~Vanzo,
  Phys.\ Rev.\  D {\bf 56}, 6475 (1997)
  [arXiv:gr-qc/9705004];
  R.~G.~Cai, J.~Y.~Ji and K.~S.~Soh,
  Phys.\ Rev.\  D {\bf 57}, 6547 (1998)
  [arXiv:gr-qc/9708063];
   D.~Klemm, V.~Moretti and L.~Vanzo,
  Phys.\ Rev.\  D {\bf 57}, 6127 (1998)
  [Erratum-ibid.\  D {\bf 60}, 109902 (1999)]
  [arXiv:gr-qc/9710123];
  D.~Birmingham,
  Class.\ Quant.\ Grav.\  {\bf 16}, 1197 (1999)
  [arXiv:hep-th/9808032];
  R.~Aros, R.~Troncoso and J.~Zanelli,
  Phys.\ Rev.\  D {\bf 63}, 084015 (2001)
  [arXiv:hep-th/0011097].
 M.~Cvetic, S.~Nojiri and S.~D.~Odintsov,
  Nucl.\ Phys.\  B {\bf 628}, 295 (2002)
  [arXiv:hep-th/0112045];
Y.~M.~Cho and I.~P.~Neupane,
  Phys.\ Rev.\  D {\bf 66}, 024044 (2002)
  [arXiv:hep-th/0202140];
I.~P.~Neupane,
  Phys.\ Rev.\  D {\bf 67}, 061501 (2003)
  [arXiv:hep-th/0212092].
\bibitem{Cai}R.~G.~Cai,
  Phys.\ Rev.\  D {\bf 65}, 084014 (2002)
  [arXiv:hep-th/0109133];
   R.~G.~Cai and Q.~Guo,
  Phys.\ Rev.\  D {\bf 69}, 104025 (2004)
  [arXiv:hep-th/0311020];
Z.~K.~Guo, N.~Ohta and T.~Torii,
  Prog.\ Theor.\ Phys.\  {\bf 120}, 581 (2008)
  [arXiv:0806.2481 [gr-qc]].
 Z.~K.~Guo, N.~Ohta and T.~Torii,
  Prog.\ Theor.\ Phys.\  {\bf 121}, 253 (2009)
  [arXiv:0811.3068 [gr-qc]];
N.~Ohta and T.~Torii,
  arXiv:0902.4072 [hep-th];

\bibitem{Cai04}R.~G.~Cai,
  Phys.\ Lett.\  B {\bf 582}, 237 (2004)
  [arXiv:hep-th/0311240].
\bibitem{CS}R.~G.~Cai and K.~S.~Soh,
  Phys.\ Rev.\  D {\bf 59}, 044013 (1999)
  [arXiv:gr-qc/9808067].

\bibitem{BTZ}M.~Banados, C.~Teitelboim and J.~Zanelli,
  Phys.\ Rev.\  D {\bf 49}, 975 (1994)
  [arXiv:gr-qc/9307033].

\bibitem{RT}T.~Regge and C.~Teitelboim,
  Annals Phys.\  {\bf 88}, 286 (1974).

\bibitem{HP}S.~W.~Hawking and D.~N.~Page,
  Commun.\ Math.\ Phys.\  {\bf 87}, 577 (1983).

\end{thebibliography}
\end{document}